\newcommand{\sun}{\odot} 

\documentclass[
    ,final            
  ]
  {aipproc}

\layoutstyle{6x9}

\begin{document}

\title{Detecting Transits in Sparsely Sampled Surveys}

\classification{97.82.Cp, 97.80.Hn}
\keywords      {Planets, Earth-Like Planets, Transits, Eclipsing Binaries, Sparsely Sampled Surveys}

\author{H. C. Ford}{
  address={Department of Physics \& Astronomy, Johns Hopkins University, Homewood Campus,
  Baltimore, MD 21218, USA}
  } 

\author{W. Bhatti}{
  address={Department of Physics \& Astronomy, Johns Hopkins University, Homewood Campus,
  Baltimore, MD 21218, USA}
  }

\author{L. Hebb}{
  address={School of Physics and Astronomy, University of St. Andrews, North Haugh, St. Andrews
  KY16 9SS, UK}
  }

\author{L. Petro}{
  address={Space Telescope Science Institute, 3700 San Martin Dr., Baltimore, MD 21218, USA}
  }

\author{M. Richmond}{
  address={Department of Physics, Rochester Institute of Technology, Rochester, NY 14623, USA}
  }

\author{J. Rogers}{
  address={Department of Physics \& Astronomy, Johns Hopkins University, Homewood Campus,
  Baltimore, MD 21218, USA}
  }
  
\begin{abstract} 
The small sizes of low mass stars in principle provide an opportunity to find Earth-like planets
and ``super Earths'' in habitable zones via transits. Large area synoptic surveys like
Pan-STARRS and LSST will observe large numbers of low mass stars, albeit with widely spaced
(sparse) time sampling relative to the planets' periods and transit durations. We present simple
analytical equations that can be used to estimate the feasibility of a survey by setting upper
limits to the number of transiting planets that will be detected.  We use Monte Carlo
simulations to find upper limits for the number of transiting planets that may be discovered in
the Pan-STARRS Medium Deep and 3$\pi$ surveys.  Our search for transiting planets and M-dwarf
eclipsing binaries in the SDSS-II supernova data is used to illustrate the problems (and
successes) in using sparsely sampled surveys.
\end{abstract}

\maketitle


\section{Planets Orbiting Low Mass Stars}

Planets transiting low mass (M $\le$ 0.6 M$_{\sun}$) stars are important for many reasons. To
$\sim$10\% accuracy, the radii of stars with M $\le$ 1 M$_{\sun}$ are $R_{star}/R_{\sun} \sim
M_{star}/M_{\sun}$. Consequently, it is easier for ground based telescopes to detect planets of
all sizes, especially Neptune-sized ice-giant planets, around M-dwarfs. Table \ref{table_1}
illustrates this point by showing the ratio of geometrical transit depths expressed as
magnitudes of Jupiter, Neptune, and the Earth transiting stars of different masses. The radii of
the stars were calculated using $R_{star}/R_{\sun} = 0.927 \times M_{star}/M_{\sun} + 0.037 \pm 0.02$,
which is a linear fit to the masses and radii of 38 low mass stars (see
\cite{Lop05,Heb06,Pon06,Bou05,Del99,Del00,Lan02}).

\begin{table}[!b]
\begin{tabular}{ccccccc}
\hline
  $\mathbf{M_{star}/M_{\sun}}$ 
& $\mathbf{R_{star}/R_{\sun}}$ 
& \textbf{Spectral Type} 
& $\mathbf{M_{i^*}/M_{z^*}}$ 
& \textbf{Jupiter} 
& \textbf{Neptune} 
& \textbf{Earth} \\
\hline
\textbf{1.00} & 1.00 & G2 & - & 0.011 & 0.001 & 0.0001\\
\textbf{0.65} & 0.64 & M0 & 8.4/7.9 & 0.028 & 0.003 & 0.0002\\
\textbf{0.45} & 0.45 & M2 & 8.9/8.3 & 0.057 & 0.005 & 0.0004\\
\textbf{0.15} & 0.18 & M4 & 10.7/9.9 & 0.447 & 0.036 & 0.0029\\
\textbf{0.092} & 0.12 & M6 & 13.2/11.9 & 1.299 & 0.076 & 0.0061\\
\textbf{0.079} & 0.11 & M8 & 14.9/13.2 & 2.118 & 0.094 & 0.0075\\
\hline
\end{tabular}
\caption{Radii, Spectral type, and geometric transit depth in magnitudes of transits for planets
  as a function of the mass of the parent star.}
\label{table_1}
\end{table}

With a limit of $\sim$2 mmag light-curve precision, transiting Neptune-sized planets are very
difficult to detect via transits around main sequence G-dwarf host stars, even in the best
ground-based photometric data. However, a Neptune-sized planet orbiting a 0.4 M$_{\sun}$ M-dwarf
has a detectable transit with a depth of 0.036 mag, and becomes more detectable as the host mass
decreases. Observing these transits around M-dwarfs will yield the radii of Neptune-like planets
and rocky planets. These radii, combined with radial velocity measurements, determine the
planets' densities, and thus provide important information about their mean
composition. Reflected and emitted light measurements, in addition to transmission spectra,
constrain the physics and composition of planetary atmospheres.

\section{Estimating the Number of Planets Transiting Low Mass Stars in a Given Survey}

We will show why large area surveys are needed to find appreciable numbers of planets transiting
low mass stars by beginning with some simple geometrical considerations. Defining the
inclination of the planet's orbit as the angle between the line of sight and the orbital plane,
the critical angle for a transit to occur is
\begin{equation}\label{eq_crit}
\sin\theta_c = (R_s + R_p)/R
\end{equation}
\noindent where $R_s$ and $R_p$ are the radii of the star and planet respectively, and R is the
radius of the circular orbit. If there are $N_p$ planet bearing stars with randomly oriented
orbits, the number of stars with planets that \emph{can} transit (as opposed to the number that
will be observed), is
\begin{equation}\label{eq_can_trans}
N_0 = N_p(R_s+R_p)/R,
\end{equation}
\noindent showing the strong selection for short period planets. The number of planets $N_I$
transiting at any given instant is $N_p$ times the fractional area of a circle with radius $R_s
+ R_p$ projected onto a sphere of radius $R$, and is given by the equation
\begin{equation}\label{eq_trans_given}
N_I = N_p \times (1/2)(1-\cos\theta_c).
\end{equation}
\noindent Equation \ref{eq_trans_given} is easily understood by thinking of the projection of
$R_s + R_p$ onto a sphere with radius R as a ``polar cap'', with the line of sight along the
axis of the pole. 

We can now estimate the number $N_t$ of planets that can possibly transit during a continuous
observation time $t_{obs}$ that is less than half the planet's period by first considering an
observation time $t_{1/4}$ that lasts for 1/4 of a period. If we suppose that all of the
transiting planets have zero inclination, then half of the planets $N_0$ can reach the polar cap
and transit, and half of this number will have an orbit toward the polar cap, and in fact
transit. Consequently,
\begin{equation}\label{eq_will_trans}
N_{t} \approx \left(N_p/4\right) (R_s + R_p)/R.
\end{equation}
\noindent Because not all transiting planets have zero inclination, Equation \ref{eq_will_trans}
is an upper limit to the number of transits that will be observed, and becomes a progressively
better estimate as the critical angle $\theta_c$ becomes smaller (i.e. as the period increases
for a fixed stellar mass). Taking into account the fraction of planets that project onto the
polar cap, an approximate upper limit for the number of planets that are transiting or will
transit during $t_{obs}$ is
\begin{equation}\label{eq_upper_lim}
N_t =
 \frac{N_p}{2}\left[\left(\frac{R_s+R_p}{R}\right)\left(\frac{\cos\theta_c-\cos\theta_t}{2}\right)
  + (1-\cos\theta_c)\right]
\end{equation}
\noindent where the angle $\theta_t$ is given by
\begin{equation}\label{eq_theta_t}
\theta_t = \theta_c + 2 \pi t_{obs}/\mathrm{Period}.
\end{equation}

We next estimate the number of stars $N(D,l,b)$ that will be in a survey area $d\Omega$ to
distance $D$ corresponding to a limiting magnitude $m_{lim}$. Although the density of stars is a
function of Galactic longitude and latitude, M-dwarfs are faint enough that, depending on
$m_{lim}$, only a few hundred parsecs will be sampled. Consequently, for a first order model, we
can assume an exponential density as a function of Galactic latitude. The number of stars is
then
\begin{equation}\label{eq_n_db1}
N(D,b) = N_0(m_1,m_2)d\Omega \int_{0}^{D} r^2 e^{-\frac{r \sin b}{r_0}} dr
\end{equation}
\noindent where $N_0(m_1,m_2)$ is the density of stars of a particular type in the galactic
plane, and $r_0$ is the scale height perpendicular to the plane. To calculate $N_0(m_1,m_2)$ we
assume a mass function of the form
\begin{equation}\label{eq_dn_dm}
\frac{dN}{dM} = 0.0187 m^{-\alpha} \mathrm{stars\ pc^{-3}}
\end{equation}
\noindent where the normalization and coefficient $\alpha = 1.3$ are taken from Reid et
al. \cite{Rei02}. The equation integrates to give
\begin{equation}\label{eq_n0_m1m2}
N_0(m_1,m_2) = \frac{0.0187}{1-\alpha} \left( m_2^{1-\alpha} - m_1^{1-\alpha}\right)
\mathrm{stars\ pc^{-3}}.
\end{equation}

We now use the preceding equations to estimate how many M-dwarfs of a given type will be in a
particular survey. As an example, consider stars with masses between 0.1375 and 0.175
M$_{\sun}$, corresponding to a mean spectral type of M4. Using Equation \ref{eq_n0_m1m2}, the
space density of M4 stars is $\sim 0.008\ \mathrm{stars\ pc^{-3}}$; if the 5$\sigma$ limiting
magnitude is $z^* = 21.0$, the limiting distance is $\sim$1700 pc, and there are $\sim$1300 M4
stars/sq-deg at $b = 15$ deg. However, we show in the next section that detection of
Neptune-sized objects with a SNR of 5 in a single observation requires decreasing the 5$\sigma$
limiting magnitude by 3.6 mags to $z^* = 17.4$, dropping the number of M4 stars to $\sim$300 per
sq-deg. If we assume 1 hour of continuous observing and that all of the planets are Neptune-like
in a $\sim$2 day orbit, the fraction of planets that can transit (Equation \ref{eq_can_trans})
is 0.05, and the fraction that will transit (Equations \ref{eq_upper_lim} and \ref{eq_theta_t})
is $6.2\times10^{-4}$. If we make the optimistic assumption that 10\% of all M4 stars have
short-period planets, the probability that a transit occurs during a 1 hour observation is
$6.2\times10^{-5}$ per planet bearing star. We must survey nearly 55 sq-deg for an hour to
detect just one transiting planet!

Surveying at low Galactic latitudes increases the number of M4 stars by an order of magnitude,
at the price of decreasing the photometric accuracy because of crowding, and greatly increasing
the number of false positives from blends of single stars with the target variable
stars. Calculations show that observing at infrared wavelengths where the M dwarfs are much
brighter helps very little from the ground because the gains from increased luminosities are
completely offset by the increases in sky background.

We look at the prospects of finding planets transiting low mass stars in sparsely sampled, wide
area surveys below.

\section{The Pan-STARRS-1 Survey}

The prototype telescope for the Panoramic Survey Telescope and Rapid Response System
(Pan-STARRS)\footnote{See: \url{http://panstarrs.ifa.hawaii.edu/public/}}, PS1, is presently
being commissioned at the Haleakala Observatories in Hawaii. Three PS1 programs are relevant for
finding transiting planets. The first, Pan-Planets \cite{Afo07}, consists of 3 campaigns of
$i$-band bright time imaging of several adjacent fields in the Galactic bulge, the Hyades, and
in Praesepe. The Galactic bulge fields will include $\sim$480,000 dwarf stars to a limiting
magnitude of $i = 17.0$; $\sim$100 short period transiting Jupiter-sized planets are expected to
be found. The host stars will be bright enough for the planets to be confirmed by radial
velocity observations. Unfortunately, the small number of M-dwarfs in these fields
($\sim$15,000; \cite{Afo07}) brighter than the limiting magnitude will lead to very few
detections of Jupiter and Neptune-sized planets around these stars.

The second of these programs, the PS1 Medium Deep Survey (MDS), will survey up to 7 fields (one
field is 7 sq-deg) every usable night, covering a total of 10 fields in a year. Each field will
be observed in g and r, i, z, and y over the course of four nights.  Assuming $65\%$ of the
nights are clear, the cycle of five filters in four nights will be repeated 35 times per year
for three years.  The $5\sigma$ limiting magnitudes for an $i$-band exposure on a single night
is given in Table \ref{table_2}.  The PS1 goal for absolute photometry precision is 0.01 mag. To
detect the 0.036 mag depth (0.033 in flux) of a Neptune transiting an M4 star in a single
observation, the photometric noise in the star must be $\le$ 0.007 mags. Assuming that
differential photometry will meet this requirement, the magnitude cutoff for the survey must be
decreased by 3.6 magnitudes. The average of the estimated number of M-dwarfs with masses $\le$
0.175 M$_{\sun}$ ($\sim$M3.5) in the MDS $i$-band and $z$-band images is $\sim 24,000$. To first
order the average of the numbers in the $r$ and $y$ filters will be comparable. The majority of
the stars will have spectral types between M3.5 and M5. Taking $\mathrm{M} =
0.1375\mathrm{M}_{\sun}$ ($\sim$M4.5) as representative of these stars and \emph{assuming all of
them have a Neptune-sized planet} with a period drawn from a uniform random distribution of
periods between 1 and 3 days, a Monte Carlo simulation of three years of observations gives the
transits listed in Table \ref{table_2}.

\begin{table}[!t]
\begin{tabular}{ccccccccc}
\hline
  \tablehead{1}{c}{c}{Survey}
& \tablehead{1}{c}{c}{Filter}
& \tablehead{1}{c}{c}{Num. Obs./\\Field/Year}
& \tablehead{1}{c}{c}{Obs.\\Sep.\\Nights}
& \tablehead{1}{c}{c}{Survey\\Limit\tablenote{The $3\sigma$ limiting magnitudes in the
3$\mathrm{\pi}$ survey have been increased by 0.75 mag based on the assumption that the two
images in a single night that are separated by 30 to 60 minutes are combined.}\\($5\sigma$)}
& \tablehead{1}{c}{c}{Survey\\Limit\\($140\sigma$)}
& \tablehead{1}{c}{c}{Num. of\\Stars}
& \tablehead{1}{c}{c}{Trans.\\Planets\tablenote{This number is derived by assuming that
\emph{$10\%$} of the stars in the Monte Carlo simulation have a short period Neptune-sized planet}}
& \tablehead{1}{c}{c}{Avg. Num.\\Trans/Star} \\
\hline
\textbf{MDS} & $r,i,z,y$ & $\sim$ 140 & 1 & 25.4 & 21.8 & $\sim$24,000 & $< $ 120 & 10.0\\
\textbf{3$\mathrm{\pi}$} & $i$ & 2 & 30 & 23.35 & 19.7 & $9.7\times10^6$ & $< $ 4,000 & 1.04\\
\textbf{3$\mathrm{\pi}$} & $z$ & 2 & 30 & 23.35 & 18.7 & $8.3\times10^6$ & $< $ 3,400 & 1.04\\
\hline
\end{tabular}
\caption{Estimated {\it upper limits} to the number of detectable planets transiting low mass
stars in the Pan-STARRS MDS and 3$\pi$ surveys. Columns 1--6 describe the assumed cadence for
the MDS and $3\pi$ surveys, the $5\sigma$ limiting magnitudes for a single exposure, and the
$140\sigma$ limits for a $5\sigma$ detection of a Neptune-sized planet transiting an M4
dwarf. The last three columns give the number of stars in the respective survey areas to the
$140\sigma$ limit, the number of planets that will transit if $10\%$ of these stars have a
Neptune-sized planet in an orbit with a period drawn from a uniform random distribution between
1 and 3 days, and the average number of transits per star.}
\label{table_2}
\end{table}

Even if we assume that 10\% of all late M-dwarfs have a short period Neptune-sized planet, the
MDS numbers are relatively small, given that only a fraction of planets that do transit during
an observation will be detected in the face of random noise, red noise, blending with eclipsing
binaries or variable stars, and other effects. The good news is that there should be $\sim$ 10
transits per short period transiting planet in the MDS survey, which will increase their
detectability.

Table \ref{table_1} shows that transiting Jupiter-sized planets are far more detectable around
the earlier and brighter M-dwarfs than Neptune-sized planets. Gaudi \cite{Gau07} concludes that
$\sim$40 such planets may be found by the MDS. This estimate, however, depends on the unknown
frequency of such higher mass planets around M-dwarfs.

The final PS1 program that we consider, the $3\pi$ steradian survey, will observe the entire sky
visible from Hawaii approximately once a week. Each field will be observed 4 times a year in
each of the five filters for 3 years (60 epochs in all filters). Each night that a field is
observed in a particular filter, the field will be observed twice with a separation of 30--60
minutes to identify moving solar system objects. We (arbitrarily) assume that the observations
two nights per year for each filter per field will be separated by 30 days, and the transits
observed in the $i$ and $z$ images are for different stars. A Monte Carlo simulation spanning
three years yields the numbers given in Table \ref{table_2}. Again, assuming that 10\% of all
low mass M-dwarfs have short period Neptune-sized planets, the $3\pi$ survey could yield $\sim$
300 such stars with 2 possible transits each.  Detecting planets which transit only twice during
widely separated short observation will be very difficult.  Unless the cadence between the {\it
\ r,i,z,y} filters is optimized for detecting transits in the $3\pi$ survey, few planets
transiting low mass stars will be detected.

Another issue that must be addressed is stellar activity in fast-rotating, fully convective low
mass dwarfs. Two recent encouraging photometric campaigns on the young active M1Ve star AU Mic
demonstrated that the star's flare activity, its 4.85 day rotation, and variability due to two
large star spots, did not prevent differential photometry accurate to a few mmag \cite{Heb07}. A
modest photometric study of late M-dwarfs could be used to assess the limits on transit
detection set by stellar activity. The next step is to make more realistic simulations that
include noise sources, specific detection algorithms, and search for optimal survey cadences.

\section{The SDSS-II Supernova Survey}

The SDSS-II Supernova Survey \cite{Fri08}, in its three years of operation, monitored an
$\sim$300 sq-deg area on the Celestial Equator known as SDSS Stripe 82 for three months per
season, with a rough cadence of one observation every two nights. We used photometric
catalogs\footnote{Available at: \url{http://www.sdss.org/drsn1/DRSN1_data_release.html}} from
the Survey to search for low mass M-dwarf binaries and transiting planets. We extracted objects
that were classified as stars with clean photometry
\footnote{\url{http://cas.sdss.org/dr5/en/help/docs/algorithm.asp}} and enforced an
upper magnitude limit of $z = 21.0$ mag. We then matched detections of objects across all three
seasons to build up an extensive light-curve catalog.

From this catalog, we selected objects that had $\ge$ 10 observations over the three seasons,
leaving us with approximately 1.1 million point sources. We used ensemble photometry techniques
\cite{Hon92} to generate differential magnitude light-curves\footnote{See M. Richmond's
implementation: \url{http://spiff.rit.edu/ensemble/}}, thus decreasing the chance of
incorrectly flagging an object as periodically variable if its light-curve was affected by the
systematic artifacts caused by changing sky conditions and/or small changes in photometric zero
points. We obtained an average photometric precision of about 30 mmag at $i = 19.0$ mag, which
effectively rules out the possibility of finding transiting planets around low mass stars, but
is comfortably within the detection regime of M-dwarf eclipsing binaries.

We used SDSS $r-i$ and $i-z$ color cuts defined by West et al. \cite{Wes05} to select
$\sim$690,000 likely M-dwarfs from our sample of 1.1 million point sources limited to $z = 21.0$
mag. To find periodic variables, we employed the Stetson variability index \cite{Ste96}; this
removed from consideration a large number of faint objects that appeared to be variable because
of photometric noise. We then inspected the differential magnitude light-curves of
color-selected likely M-dwarfs that survived this selection process, looking for multi-band
events that were eclipse-like.

\begin{figure}[!t]\label{fig_lcs}
  \includegraphics[width=0.475\textwidth]{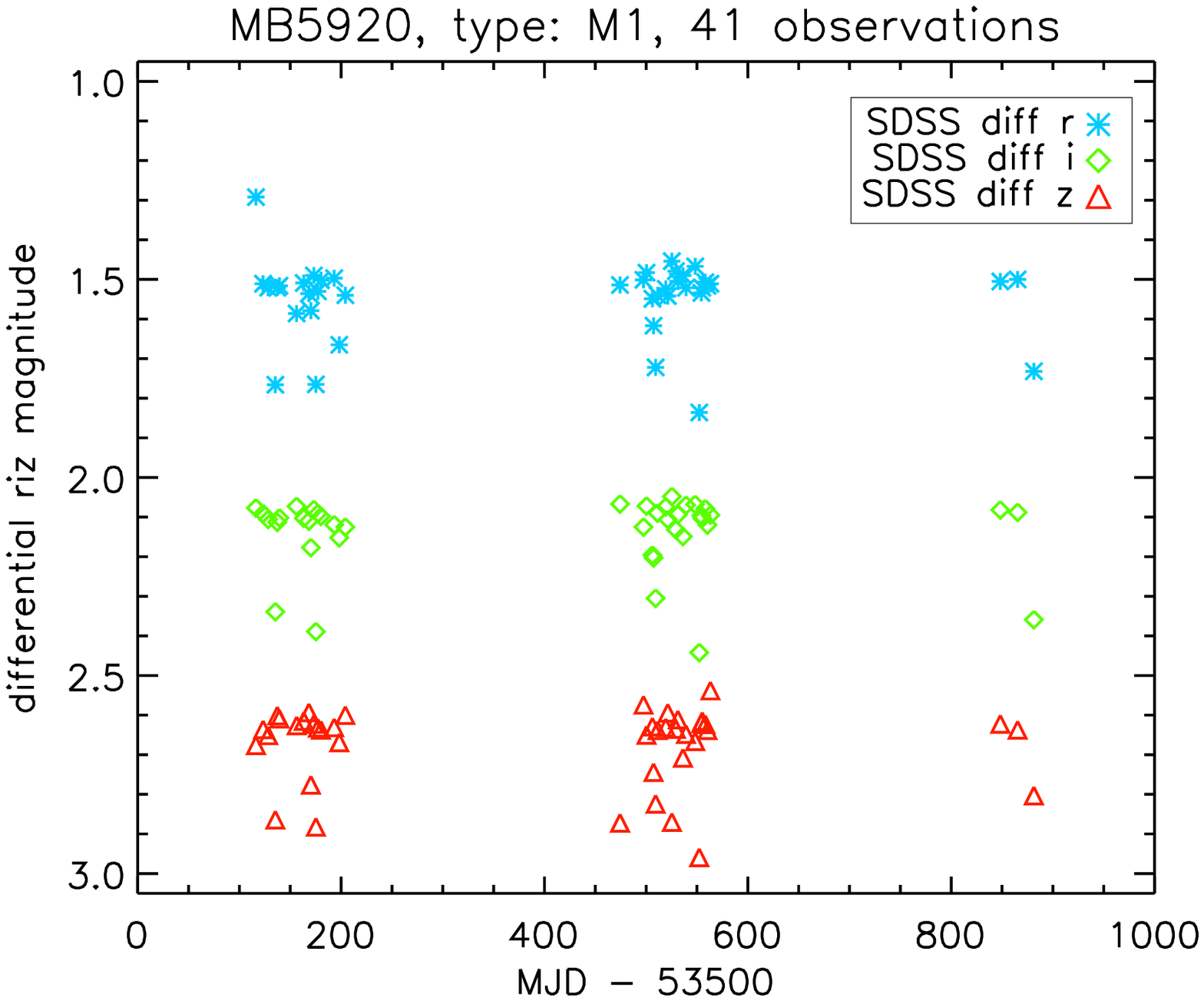}
  \includegraphics[width=0.475\textwidth]{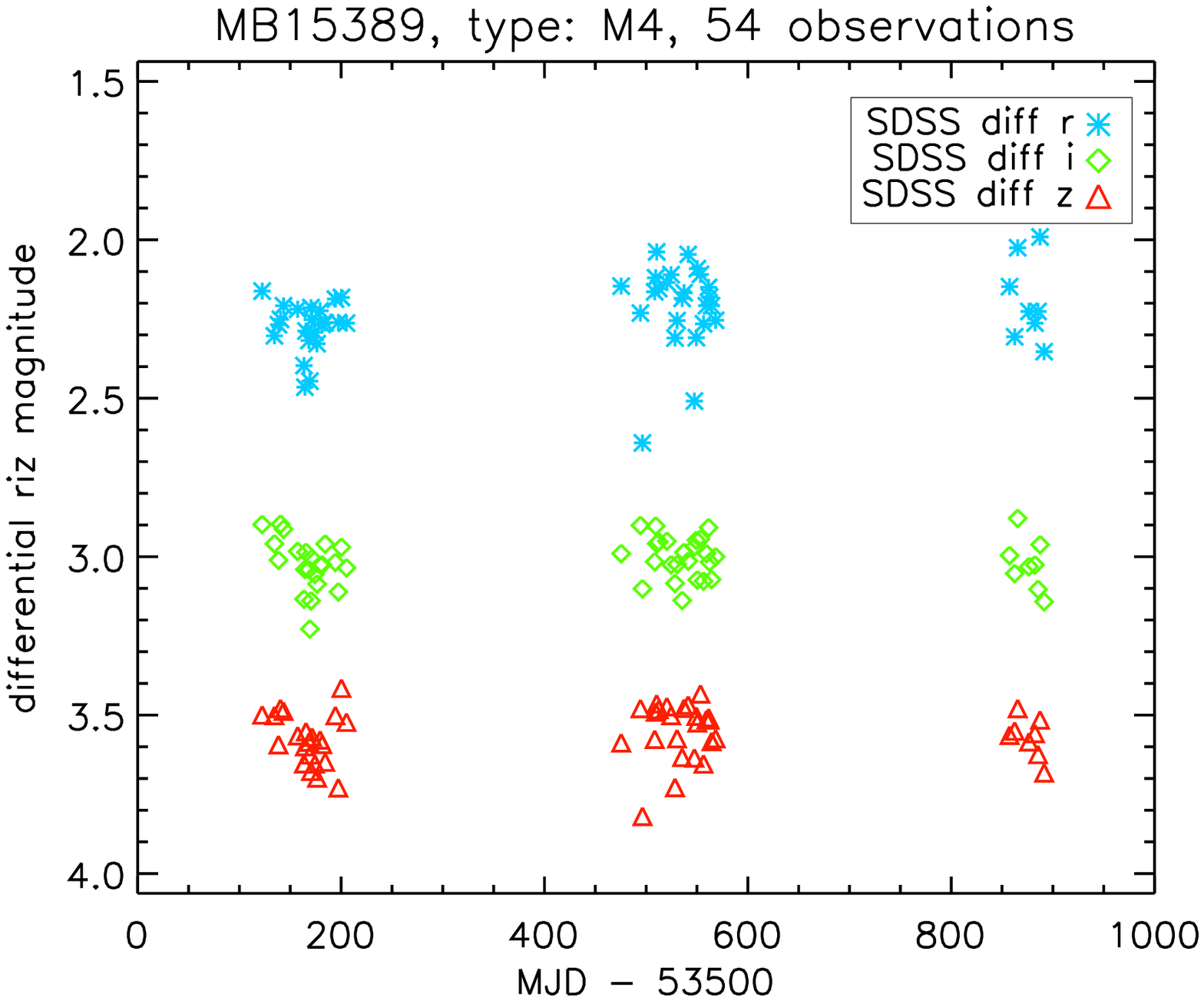} 
\caption{Differential photometry $riz$ light-curves for two M-dwarf eclipsing binary candidates
  from the SDSS-II Supernova Survey: MB5920 (left panel) and MB15389 (right panel). The
  differential magnitudes have been shifted on the y-axis for clarity. Both candidate names are
  internal designations.}
\end{figure}

A rough estimate of the total number $N_{det}$ of eclipsing binary M-dwarf systems we expect to
find is given by
\begin{equation}\label{eq_n_det}
N_{det} = N_{*} \times f_{M} \times f_{bin}(P) \times f_{trans} \times f_{thresh}. 
\end{equation}
\noindent $N_{*}$ is the total number of stars in our sample, $f_{M}$ is the fraction of
M-dwarfs seen in the sample, $f_{bin}(P)$ is the M-dwarf binary fraction as a function of
period, $f_{trans}$ is the fraction of M-dwarf binary systems that can be observed by our
survey, and $f_{thresh}$ is the fraction of M-dwarf eclipsing binary systems that will be
detected given our survey's SNR and throughput limits. Using the numbers from our sample for
$N_{*} \simeq 1.1\times10^6$, $f_{M} \simeq 0.60$, and conservative estimates for
$f_{bin}(\mathrm{2.5-7\ days}) \simeq 0.01$, $f_{trans} \simeq 0.05$, and $f_{thresh} \simeq
0.05$, we estimate that we may be able to find up to 17 M-dwarf eclipsing binaries.

To date, we have processed $\sim$30\% of our sample and have found 12 candidates. We present the
light-curves of two such objects in Figure \ref{fig_lcs}. We are currently testing various
period-finding techniques, including BLS \cite{Kov02}, and AoV \cite{Sch96}, to determine the
likely periods of these systems. This will allow us to generate ephemerides for photometric
monitoring, and enable eventual radial velocity followup to confirm whether they are indeed
eclipsing binary systems of low mass M-dwarfs.


\begin{theacknowledgments}

This research was supported by NASA grant NAG5-7697, and was based in part on data from the
Sloan Digital Sky Survey (\url{http://www.sdss.org}). We thank Ani Thakar at JHU and Brian Yanny
at FNAL for helping us access the SDSS-II Supernova Survey data.

\end{theacknowledgments}


\end{document}